# Unusual magnetic-field dependence of partially frustrated triangular ordering in manganese tricyanomethanide

R. Feyerherm[1], A. Loose[1], J. L. Manson[2]

[1] *Hahn-Meitner-Institute and Berlin Neutron Scattering Center, 14109 Berlin, Germany*
[2] *Solid State Division, Oak Ridge National Laboratory, Oak Ridge, TN 37831-6430, U.S.A.*

Full address of corresponding author:

Dr. Ralf Feyerherm
Hahn-Meitner-Institute
Department SF2
Glienicker Str. 100
14109 Berlin
GERMANY
Phone: ++49 30 8062 3082
Fax:    ++49 30 8062 3172
E-mail: Feyerherm@hmi.de

**Short title:** Partially frustrated triangular ordering in $Mn[C(CN)_3]_2$

**Classification Nos.:** 75.25, 75.40.C, 75.50.E

**Abstract.**   Manganese tricyanomethanide, $Mn[C(CN)_3]_2$, crystallizes in an orthorhombic lattice consisting of two interpenetrating three-dimensional rutile-like networks. In each network, the tridentate $C(CN)_3^-$ anion gives rise to superexchange interactions between the $Mn^{2+}$ ions ($S = 5/2$) that can be mapped onto the "row model" for partially frustrated triangular magnets. We present heat capacity measurements that reveal a phase transition at $T_N = 1.18$ K, indicative of magnetic ordering. From magnetic-field dependent heat capacity data a saturation field $H_{sat} = 42$ kOe is estimated. The zero-field magnetically ordered structure was solved from neutron powder diffraction data taken between 0.04 and 1.2 K. It consists of an incommensurate spiral with a temperature independent propagation vector $\boldsymbol{Q} = [2Q\ 0\ 0] = [\pm 0.622\ 0\ 0]$, where different signs relate to the two different networks. This corresponds to $[\pm 0.311\ \pm 0.311\ 0]$ in a quasi-hexagonal representation. The ordered moment $\mu = 3.3\ \mu_B$ is about 2/3 of the full $Mn^{2+}$ moment. From the values of $T_N$ and $Q$, the exchange parameters $J/k_B = 0.15$ K and $J'/J = 0.749$ are estimated. The magnetic-field dependence of the intensity of the $(2Q\ 0\ 0)$ Bragg reflection, measured for external fields $\boldsymbol{H}\ //\ \boldsymbol{Q}$, indicates the presence of three different magnetic phases. We associate them with the incommensurate spiral ($H < 13.5$ kOe), an intermediate "up-up-down" phase (13.5 kOe $< H <$ 16 kOe), and the "2-1" spin-flop like magnetic structure ($H > 16$ kOe) proposed for related compounds. For increasing fields, $Q$ continuously approaches the value 1/3, corresponding to the commensurate magnetic structure of the fully frustrated triangular lattice. This value is reached at $H^* = 19$ kOe. At this point, the field-dependence reverses and $Q$ adopts a value of 0.327 at 26 kOe, the highest field applied in the experiment. Except for $H^*$, the magnetic ordering is incommensurate in all three field dependent magnetic phases of $Mn[C(CN)_3]_2$.

## 1. Introduction

A number of organic ligands have been shown to be able to provide magnetic superexchange pathways between transition metal ions leading to a class of molecule based magnetic materials [1]. Bidentate ligands result in a variety of interesting structural topologies and magnetic ground states. Recently, we have investigated, e.g., the quasi one-dimensional (1D) Heisenberg antiferromagnets Cu(pym)(NO$_3$)$_2$(H$_2$O)$_2$ [2] and Cu(pym)(NO$_3$)$_2$ [3] with pym = pyrimidine, the magnetic ordering in the two-dimensional (2D) network systems Co(ox)(bpy) and Co/NiCl$_2$(bpy) with ox = oxalate (C$_2$O$_4^{2-}$), bpy = 4,4'-bipyridine [4,5] and the three-dimensional (3D) network Mn(dca)$_2$(pyz) with dca = dicyanamide, [N(CN)$_2$]$^-$, and pyz = pyrazine [6].

Tridentate ligands are expected to add another flavor to magnetism in molecule-based compounds, because they may give rise to triangular topologies and therefore to magnetic frustration. The tricyanomethanide anion, C(CN)$_3^-$, has three-fold symmetry and therefore appears to be an ideal candidate. Recently, the synthesis and magnetic properties of the series M[C(CN)$_3$]$_2$, with M = transition metal, has been reported [7-11]. As described in detail below, these compounds may be regarded as examples for partially frustrated triangular lattices represented by the "row model" introduced by Kawamura [12] and Zhang *et al.* [13]. The study of such partially frustrated systems is of a fundamental interest, because they do not simply correspond to an intermediate between unfrustrated and frustrated magnetism but show novel physical phenomena absent in the two limiting cases [14].

To date, only in the case of Cr[C(CN)$_3$]$_2$, has the magnetic ordering been studied by neutron diffraction [10]. Although frustration effects appear to reduce the ordering temperature in this compound, partial lifting of frustration results in a rather conventional magnetically ordered structure below $T_N = 6.1$ K, involving doublings of the unit cell along two crystal axes and a large ordered moment of 4.7 $\mu_B$. The critical behavior of Cr[C(CN)$_3$]$_2$ is consistent with the 2D Ising model. No ordering above 2 K was observed in V[C(CN)$_3$]$_2$ [10] as well as the Cu, Ni and Co compounds [9,11]. Here we describe heat capacity and neutron diffraction studies of the magnetic ordering below 1.18 K in Mn[C(CN)$_3$]$_2$. A previous report of an antiferromagnetic transition for Mn[C(CN)$_3$]$_2$ at $T_N = 5$ K [9] could not be confirmed by the present work.

Like the other members of the series, Mn[C(CN)$_3$]$_2$ crystallizes in an orthorhombic lattice, space group *Pmna*. Its lattice parameters are $a = 7.679$ Å, $b = 5.384$ Å, $c = 10.623$ Å at ambient temperatures [8]. The crystal structure is depicted in Figure 1. There are two Mn per unit cell belonging to two identical interpenetrating three-dimensional rutile-like networks. One network is related to the second by a full unit cell translation along *b*. The tridentate C(CN)$_3^-$ anion gives rise to superexchange interactions between the Mn ions. Along *a*, neighboring Mn are linked by two C(CN)$_3^-$, while along the other directions the Mn are linked by only one C(CN)$_3^-$. The structure suggests that the magnetic coupling within a single network can be mapped onto the "row model" for partially frustrated triangular magnets, with the stronger magnetic coupling along *a* (see Figure 4). Each network consists of 2D triangular sheets that intersect at rows parallel to *a*. Since the smallest *inter*network Mn-Mn spacing (5.384 Å) is significantly smaller that the *intra*network spacing (7.679 Å), additional magnetic interactions *between* the two networks, presumably through dipolar coupling, cannot be neglected.

In the following we will show that Mn[C(CN)$_3$]$_2$ exhibits magnetic ordering below 1.18 K. The zero-field magnetically ordered structure is an incommensurate spiral as expected for the row model. The magnetic-field dependent behavior, however, appears to be quite unusual. Our results suggest that there are three different field-dependent magnetic phases that we associate with the incommensurate spiral, an intermediate "up-up-down" phase around $H_{sat}/3$, and the "2-1" spin-flop like magnetic structure proposed for related compounds. The incommensurate propagation vector varies with increasing field, reaching the value consistent with the *commensurate* structure of ideal triangular lattices at one specific field value $H^*$ but returning to incommensurate values for higher fields. Except for $H^*$, the magnetic ordering is incommensurate in all three field dependent magnetic phases of Mn[C(CN)$_3$]$_2$.

## 2. Experimental

A microcrystalline, bright sand-colored sample of Mn[C(CN)$_3$]$_2$ resulted from the fast precipitation from an aqueous solution of Na[C(CN)$_3$]$_2$ and Mn(NO$_3$)$_2$. The sample was filtered and washed with H$_2$O. The quality of the microcystalline sample was checked by x-ray and neutron powder diffraction which were consistent with the published crystal structure [8] and exhibited no secondary phase down to the 2% level.

Heat capacity measurements were conducted using a Quantum Design PPMS (Physical Properties Measurement System) equipped with the Heat Capacity System operating on a $^3$He insert. A powder sample weighing 4.2 mg was pressed into a disc of nominally 4 mm in diameter. Because the sample is a poor thermal conductor it was necessary to make the disc very thin (~0.3 mm). The pellet was adhered to the addenda using Apiezon N grease. Because of the rather large heat capacity, with associated calorimeter time constants in excess of $1 \times 10^3$ s, a quasi-adiabatic technique was used. As a result the heat capacity values were primarily extracted from the heating portion of the calorimeter response.

Magnetic-field dependent heat capacity measurements were carried out using an Oxford Instrument 9 T MagLab calorimeter with $^3$He refrigerator and a very small amount of powder sample ($\approx 100 \mu g$) attached to an equally small amount of grease containing alumina powder (Wakefield 120-2). From these data, only the peak positions of the heat capacity curves were determined.

Neutron powder diffraction measurements were performed using the instruments E6 and E9 at the Berlin Neutron Scattering Center (BENSC). Neutron wavelengths of 2.448 Å and 1.7964 Å, respectively, were used. The instrument E6 provides a high neutron flux and medium resolution, is equipped with a 20°-multichannel detector, and covers a range of scattering angles up to about 100°. In contrast, the instrument E9 is a low-flux fine-resolution powder diffractometer with an extended $2\theta$-range up to 160°. Therefore, the former was used for the study of *magnetic* Bragg reflections, while the latter was used for the above mentioned check of the crystal structure. Magnetic-field dependent low-temperature measurements were carried out by using a $^3$He/$^4$He-dilution stick cryostat mounted to a horizontal-field superconducting magnet. The maximum field was limited to below 3 T because of strong forces between the magnet and magnetic parts of the instrument. The Rietveld refinement of the diffraction data was carried out using the WINPLOTR/FULLPROF package [15].

## 3. Results

From low-temperature heat capacity measurements, it was possible to determine the magnetic ordering temperature. A sharp $\lambda$-anomaly was observed at 1.18 K that is ascribed to 3D ordering of the Mn$^{2+}$ moments, Figure 2a. Because the magnetic ordering occurs at such low temperatures, the phonon contribution to the total heat capacity is negligible. Hence the magnetic entropy, $\Delta S_{mag}$, can be easily determined. Between 0.5 K and $T_N$ only 42% of $R\ln S(S+1) = 14.88$ J mol$^{-1}$K$^{-1}$ is released. This entropy reduction suggests that only a fraction of the Mn$^{2+}$ moments are actually ordered.

Figure 2b shows the *H-T* phase diagram determined from the position of the heat capacity anomaly in various applied magnetic fields as measured in an independent experiment. The curve was fitted with the empirical formula $H(T) = H(0)(1-T/T_N)^\alpha)^\beta$ which has no physical background but allows for the estimation of the saturation field $H_{sat} = H(0)$. We obtain $H_{sat} = 42(2)$ kOe ($T_N = 1.15$ K, $\alpha = 1.77$, $\beta = 0.386$).

Figure 3 shows the difference of zero-field neutron powder diffractograms taken at 0.04 and 1.2 K, i.e., far below and above $T_N$. It reflects the Bragg scattering of neutrons from the magnetically ordered lattice. Several magnetic Bragg reflections can be identified. On the basis of the crystallographic unit cell they are indexed (h k l) with $k = n \pm 0.6222$ and $l = 0$ or $l = n \pm ½$ ($n = 0, 1, 2, ...$), pointing to a complex magnetic ordering involving an incommensurate modulation along $a$ and an independent doubling of the unit cell along $b$. In the following we will employ a modified unit cell with $b' = 2b$, which allows for an independent description of the two networks. In this modified cell, the indexing of the magnetic Bragg reflections is simplified and only one magnetic propagation vector is necessary to describe the magnetically ordered structure.

From the close relation of the crystal structure to the row model it is expected that the incommensurate ordering corresponds to a spiral propagating along $a$ for both networks. Assuming identical magnitudes of the ordered magnetic moment at all Mn sites, there are only a few parameters that need to be determined by a Rietveld refinement of the difference diffractogram. The result of the refinement is shown in Figure 3 as solid line through the data points. The model gives an excellent fit of the experimental data ($R_p = 0.050$, $R_{wp} = 0.061$). The only variables are

- $Q = \pm 0.3111(5)$, corresponding to a rotation angle of $\alpha = \pm 112.0(2)°$ between $bc$ sheets separated by $a/2$, where the +/- signs relate to the two different networks,
- the relative orientation of the moments in the two networks at $x = 0$, putting the first spin parallel to $a$ : $\phi = 180°$ (phase shift),
- the plane in which the moments rotate: the $ab$ plane, and
- the magnitude of the ordered magnetic moment: $\mu = 3.34(5)\ \mu_B$.

All other parameters (such as lattice constants and line widths) are fixed by a fit to the powder diffractogram taken at 1.2 K (not shown). The corresponding magnetic structure is depicted in Figure 4. It is interesting to note that the moments of the two independent networks counter-rotate (left-handed vs. right-handed spiral). The alternative model with equal-handed spirals was tested extensively but did not result in any satisfactory refinement of the neutron diffraction data.

The temperature dependence of the intensity of the ($2Q$ 0 0) Bragg reflection is shown as inset in Figure 3. From these data, we derive a value of $T_N = 1.12$ K for our powder sample, slightly lower than observed in the specific heat measurements. The data shown no indications for a temperature dependence of $Q$ in zero magnetic field.

Magnetic field dependent measurements were carried out at 0.04 K applying the field parallel to the scattering vector for the ($2Q$ 0 0) Bragg reflection to within 2°. Only those grains for which $\mathbf{H} \parallel \mathbf{Q} \parallel a$ contribute to the measured intensity of this specific reflection. The corresponding diffraction data are shown in Figure 5. The ($2Q$ 0 0) peak is observed to shift with increasing magnetic field, while at the same time its intensity exhibits a pronounced field dependence. Also the other reflections exhibit a field-dependence. For these reflections $\mathbf{H}$ is roughly parallel to the corresponding scattering vectors due to the specific experimental geometry. In the following, we will focus only on the characteristics of the ($2Q$ 0 0) reflection.

The field dependence of $Q$ for $\mathbf{H} \parallel \mathbf{Q}$ is shown in the upper portion of Figure 6. For increasing fields, $Q$ continuously approaches the value 1/3 corresponding to the commensurate magnetic structure of the fully frustrated triangular lattice. This value is reached at a characteristic field $H^* = 19$ kOe. At this point, the field-dependence reverses. At 26 kOe, the highest field applied in the experiment, $Q$ adopts a value of 0.327.

The field-dependence of the intensity of the (2Q 0 0) reflection is shown in the lower portion of Figure 6. With increasing field, it exhibits a steep drop above 10 kOe and is unobservable for 13.5 kOe < H < 16 kOe. For higher fields it increases again steeply. This suggests the existence of three different magnetic phases. It is interesting to note that the field-dependence of Q below 13.5 kOe can be extrapolated through the field region of unobservable (2Q 0 0) reflection and therefore Q appears to vary continuously through this field region.

## 4. Discussion

*Zero field magnetic structure*

From the crystal structure it was anticipated that the magnetic interactions in $Mn[C(CN)_3]_2$ can be mapped onto the row model. Neglecting inter-network interactions, the corresponding Hamiltonian is

$$H = -J \sum_{<i,j>} \mathbf{S}_i \cdot \mathbf{S}_j - J' \sum_{<k,l>} \mathbf{S}_k \cdot \mathbf{S}_l - \mu_B g \sum_m \mathbf{H} \cdot \mathbf{S}_m + D \sum_n (\mathbf{S}_n^z)^2 \qquad (1)$$

where the first and second sums are over the intra- and interrow pairs of nearest neighbors, respectively, the third term is the Zeeman interaction and the last term is the anisotropy term. Due to the specific structure of $Mn[C(CN)_3]_2$, within a single network no additional superexchange pathway is present. The distance between moments along "chains" of Mn along *c* is 10.6 Å and therefore much too large for any exchange interaction. The ferromagnetic alignment of these moments can be regarded as a natural consequence of the exchange interaction through the $C(CN)_3^{2-}$. Therefore, no additional exchange parameter is necessary in the Hamiltonian of a single network and it is identical to a quasi-2D model.

We showed above that the zero-field magnetically ordered structure is an incommensurate spiral as expected for this model. Mean field theory [13] relates the values of $T_N$ and $Q$ to the exchange parameters $J$ for *intra*row and $J'$ for *inter*row interactions by the following two equations:

$$\cos(2\pi Q) = -J'/2J, \qquad (2)$$

$$k_B T_N = JS^2(1 + 0.5(J'/J)^2). \qquad (3)$$

From (2) we obtain $J'/J = 0.749$, and with this value from (3) $J/k_B = 0.15$ K.

A weak easy-plane anisotropy $D > 0$ is apparently present in $Mn[C(CN)_3]_2$, favoring the alignment of moments within the *ab* plane. The question of an additional magnetic interaction acting between the two networks is related to the phase shift defined above. Dipolar field calculations indicate that nearest-neighbor dipolar fields amount to 0.8 kG (for 3.3 $\mu_B$ ordered moment) and favor a phase shift of 180° together with counter-rotating moments as observed in the experiment. In the case of all moments rotating in the same sense, a phase shift of zero would be favored by dipolar fields, however, with an energy gain of only 1/3 compared to counter-rotating moments. Therefore, dipolar fields may be responsible for defining the relative orientation of the moments in the two networks.

The observed magnitude of the ordered moment is 68% of the full moment of $Mn^{2+}$ ions (5$\mu_B$). Due to the partial release of frustration in $Mn[C(CN)_3]_2$, the moment reduction therefore is less pronounced than in a fully frustrated triangular lattice, where a moment reduction of roughly 50% is estimated from theory [16,17]. The remaining fraction of the full moment is still fluctuating below $T_N$. The moment reduction corresponds well to the observed entropy reduction.

Only a few other triangular lattices with incommensurate magnetic ordering have been reported in the literature and may serve for comparison. $AgCrO_2$ is a quasi-2D system with $Q = 0.327$ [18]. *Ferro*magnetically stacked triangular lattices are represented by $RbCuCl_3$ with $Q = 0.2993$ [19], and $CsFeCl_3$ [20] which has a field-induced incommensurate phase with $Q = 0.322$ at 5 T but $Q = 1/3$ at higher fields. RbFeCl3 has an ideal triangular structure but a complex field and temperature dependent phase diagram with incommensurate and commensurate phases [21]. Examples for *antiferro*magnetically stacked triangular magnets with incommensurate ordering are $RbMnBr_3$ [22,23], which has complex magnetic ordering involving a lock-in transition to an eight times enlarged magnetic unit cell in applied magnetic fields, and $Cs_2CuCl_4$ with $Q = 0.47$ in zero field [24], hence a small ratio $J'/J = 0.19$. However, the physics of antiferromagnetically stacked triangular lattices is qualitatively different from the quasi-2D case and therefore the latter examples will not be discussed further.

*Field dependence*

The fact that the exchange interaction is small in $Mn[C(CN)_3]_2$ allows to influence the magnetic ground state by application of moderate magnetic fields. In the present case we observe a pronounced field dependence of the intensity of the (2$Q$ 0 0) reflection and of the value of $Q$. Only in a part of the above mentioned related compounds the magnetic field dependence of the magnetic structure has been investigated and none of them shows a behavior comparable to $Mn[C(CN)_3]_2$.

In order to understand the field-dependent behavior of $Mn[C(CN)_3]_2$ we recapitulate the field dependence of magnetic ordering in ideal triangular lattices with ferromagnetic stacking which exhibit the ideal commensurate 120° structure in the ground state. The only such compounds that were studied in detail are $RbFeCl_3$, already mentioned above, and $CsCuCl_3$. In these compounds field-induced transitions from the commensurate 120° spiral structure via a collinear up-up-down ("uud") phase to a spin-flop-like "2-1" phase has been established [25,26] which is also supported by theory [27]. The corresponding spin-arrangements are shown in Figure 7. In the collinear phase the moments are aligned parallel to the external field, while in the 2-1 state they are canted.

The question of how these transitions are modified in the case of distorted triangular lattices like the row-model has not been discussed before. Our field-dependent data are the first that address this question.

We suggest the following scenario for $Mn[C(CN)_3]_2$: the incommensurate spiral and a structure similar to the "2-1" phase are realized in the field regions $H < 13.5$ kOe and $H > 16$ kOe, respectively. These phases are separated by an intermediate phase. The analogy with the related compounds $RbFeCl_3$ and $CsCuCl_3$ suggests that this phase may be the "uud" phase. The associated alignment of the moments parallel $a$ would naturally lead to zero intensity of the (2$Q$ 0 0) reflection as observed in the experiment. Without fluctuations, the "uud" phase would exist only at a specific field value $H_{uud} = H_{sat}/3 = 14$ kOe. This value is in excellent agreement with the observed phase boundary at 13.5 kOe. It is interesting to note at this point that the value of $H_{sat}$ determined above by heat capacity measurements is in perfect agreement with the value calculated from the equation [28]

$$H_{sat} = 2S\{J(1-\cos 4\pi Q) + 2J'(1-\cos 2\pi Q)\}/\mu_B = 42 \text{ kOe} \qquad (4)$$

using $J = 0.15$ K and $Q = 0.311$.

The present data suggest that the "uud" state is stable in the extended field region 13.5 kOe $< H <$ 16 kOe. Theory, however, predicts that at low temperatures the "uud" phase is stable over a finite field range only in the presence of quantum fluctuations [27]. These are supposed to be small in $Mn[C(CN)_3]_2$, because it has a large spin value and the frustration is partially lifted. Therefore, the presence of this phase in an extended field region is rather surprising and may be related to a weak in-plane anisotropy.

At the transition to the "2-1" state, the intensity of the (2Q 0 0) reflection increases steeply, because the moment are not any more directed parallel to *a*. Surprisingly, our data suggest that, except for $H^*$, the magnetic ordering is *incommensurate in all three field dependent magnetic phases* of $Mn[C(CN)_3]_2$, including the "uud" and "2-1" states. For the "uud" state and the "2-1" phase, this may be interpreted as a slow modulation along *a*, with a repeat distance of $(1 - 3Q)^{-1}$ "uud"/"2-1" units, presumably by regular spin slips. This would distinguish the present case of a distorted triangular lattice from the ideal one. At present we have neither an explanation for this behavior nor for the peculiar field-dependence of *Q* including the fact that $Q \leq 1/3$ at all field values.

Recently, high-field magnetization and ESR studies of the distorted triangular magnet $RbCuCl_3$ revealed a phase transition at $H_c \approx H_{sat}/3 = 212$ kOe for $H \perp c$ and $T = 1.3$ K [28], which we attribute to a transition to the "uud" phase similar to the one observed in the present experiment on $Mn[C(CN)_3]_2$. Unfortunately, $H_c$ is too large for present neutron scattering instruments and therefore this assertion cannot be tested. An intermediate phase between the incommensurate spiral and the paramagnetic phase was also inferred from high-field specific heat measurements close to $T_N \approx 19$ K in this compound [29].

**Conclusion**

We have shown that $Mn[C(CN)_3]_2$ is a realization of the "row model" for partially frustrated triangular magnets and exhibits magnetic ordering below $T_N = 1.18$ K. The zero-field ordered magnetic structure consist of an incommensurate spiral, as expected for the case of partially lifted frustration. From the observed propagation vector, a ratio $J'/J = 0.749$ is estimated.

The magnetic-field dependence of the intensity of the (2Q 0 0) Bragg reflection for $H \parallel Q$ indicates the presence of three different magnetic-field dependent phases. We suggest that these phases are similar to the commensurate 120°, the "uud", and the "2-1" phases observed in related ideal triangular lattices with ferromagnetic stacking. However, in the case of $Mn[C(CN)_3]_2$ these phases appear to be incommensurate for all fields but a specific field value $H^*$. In order to verify this scenario, single crystal studies of $Mn[C(CN)_3]_2$ are necessary. Theoretical studies are needed to check whether the postulated modulation of the moment magnitudes in the up-up-down and 2-1 states can be related to the distortion of the triangular lattice, i.e., to the fact that $J \neq J'$.

Further studies on $Mn[C(CN)_3]_2$ are foreseen, including heat capacity [30] and magnetization measurements at low temperatures and in high magnetic fields, as well as additional neutron diffraction studies.

**Acknowledgments**

Work at the Oak Ridge National Laboratory was supported by the U. S. Department of Energy under contract No. DE-AC05-00OR22725 which is managed by UT-Battelle, LLC. We acknowledge valuable discussions with U. Schotte, N. Stuesser, M. Reehuis, and M. Zhitomirsky. We thank S. Kausche, P. Smeibidl, and D. Toebbens for experimental support. Special thanks go to Dr. R. Black and D. Polansic (Quantum Design) for the zero-field and to M. Meißner and T. Polinski for the field-dependent heat capacity measurements.

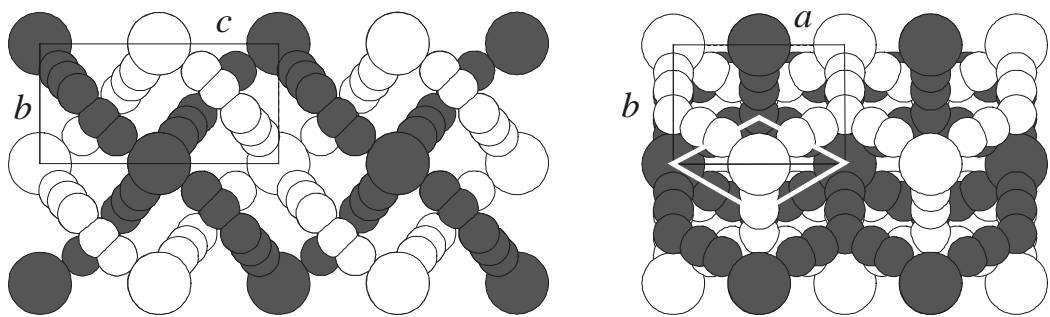

**Figure 1.** Crystal structure of Mn[C(CN)$_3$]$_2$ viewed along the a (left) and c axes (right), emphasizing the two interpenetrating networks. The Mn are depicted as big spheres, the smaller spheres are C and N atoms. Bright and dark atoms belong to the two different networks which are related by a full unit cell translation along *b*. Each network consists of 2D triangular sheets that intersect at rows parallel to *a* (see left). Along the a direction neighboring Mn are linked by two C(CN)$_3^-$ molecules as marked for one of the networks by the white diamond (right).

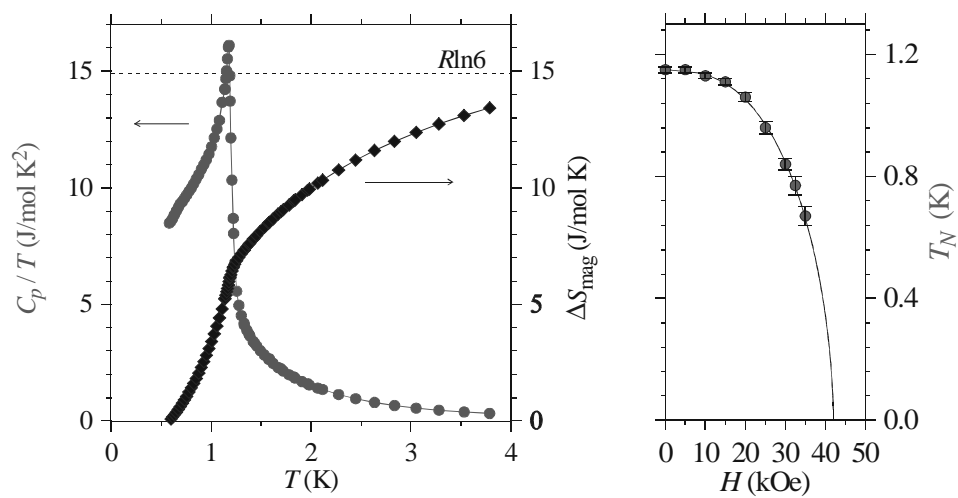

**Figure 2.** (a) Low temperature heat capacity for Mn[C(CN)$_3$]$_2$ showing the λ-anomaly associated with the long-range magnetic ordering of the Mn$^{2+}$ moments and the magnetic entropy, $\Delta S_{mag}$, for Mn[C(CN)$_3$]$_2$. The solid lines are guides to the eye only, while the dashed line delineates the magnetic entropy for $S = 5/2$. (b) Magnetic-field dependence of the position of the heat capacity anomaly, measured in an independent experiment. The solid line in the figure is a fit to an empirical formula (see text).

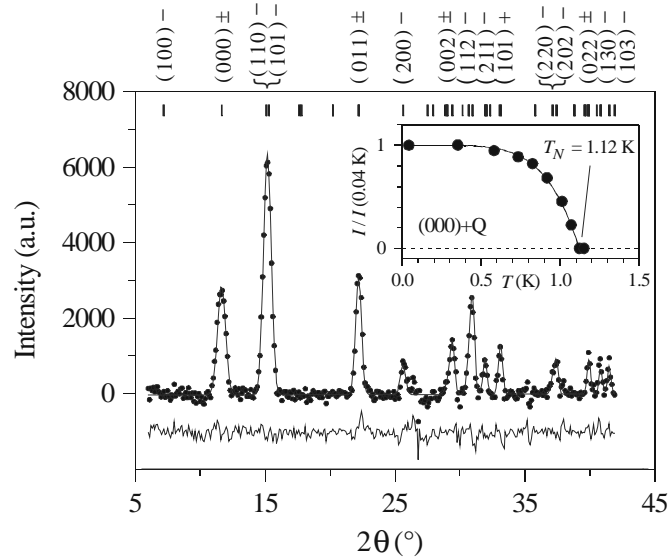

**Figure 3.** Difference of neutron powder diffractograms taken at 0.04 and 1.2 K in zero magnetic field (data points) showing the magnetic contribution to the Bragg scattering. The Rietveld refinement is shown as solid line through the data and the residual at the bottom. Magnetic Bragg reflections are indexed on the basis of a modified unit cell with $b'=2b$. Inset: temperature dependence of the intensity of the $(2Q\ 0\ 0)$ reflection. The solid line in the inset is a guide to the eye.

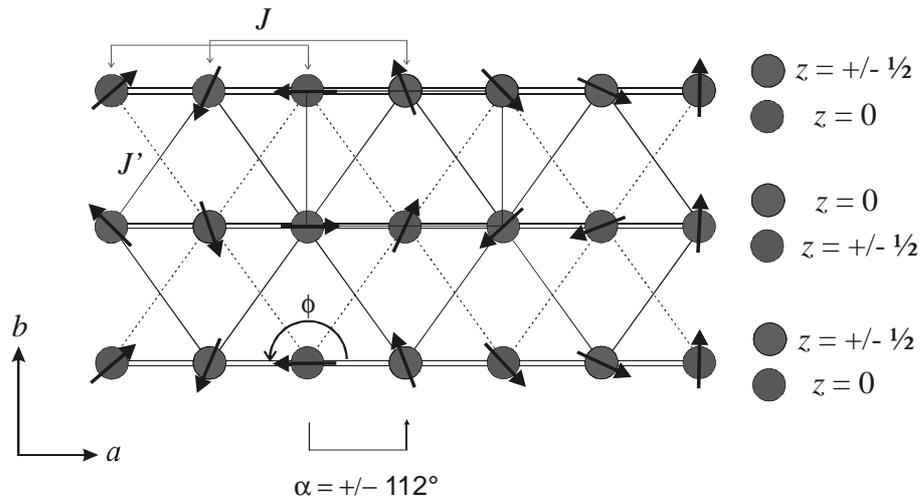

**Figure 4.** Zero-field magnetic structure of $Mn[C(CN)_3]_2$ as determined from neutron powder diffraction at 0.04 K. The view is along $c$. The crystallographic unit cell is marked. The magnetic topology of the row model is depicted as double line for magnetic coupling $J$ within rows along $a$ and single lines for coupling $J'$ between rows. Note that the moments of the two independent networks counter-rotate and have a phase shift $\phi=180°$ as marked.

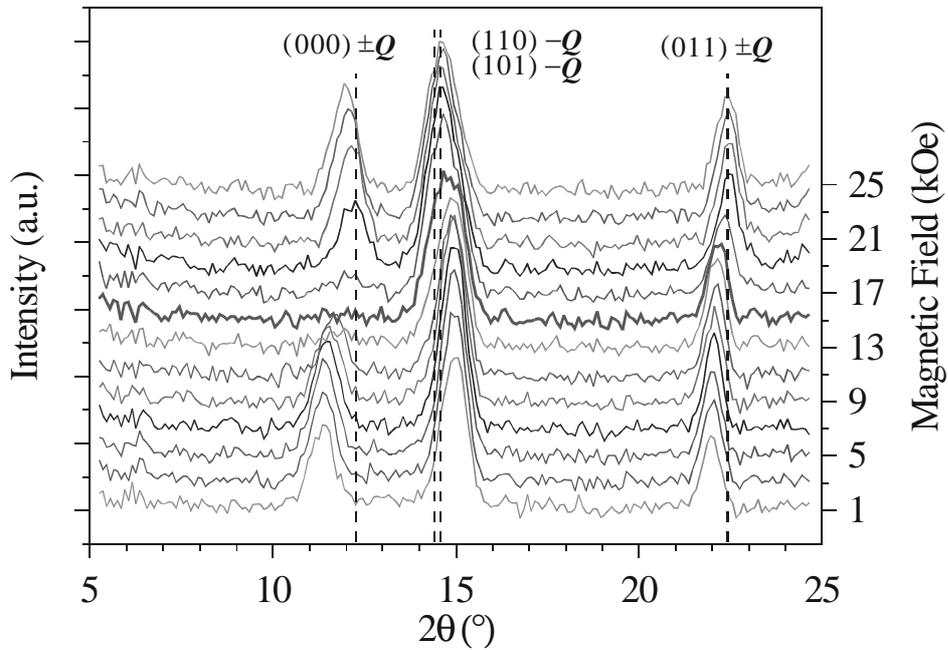

**Figure 5.** Magnetic-field dependent neutron diffraction data for Mn[C(CN)$_3$]$_2$ at $T = 0.04$ K with a horizontal field applied parallel to the scattering vector for the ($2Q$ 0 0) reflection. The data taken at 15 kOe are highlighted. The spectrum taken at 1.2 K and $H = 0$ is subtracted hence the data show only the magnetic contribution. Dashed lines indicate line positions expected for $Q = 1/3$.

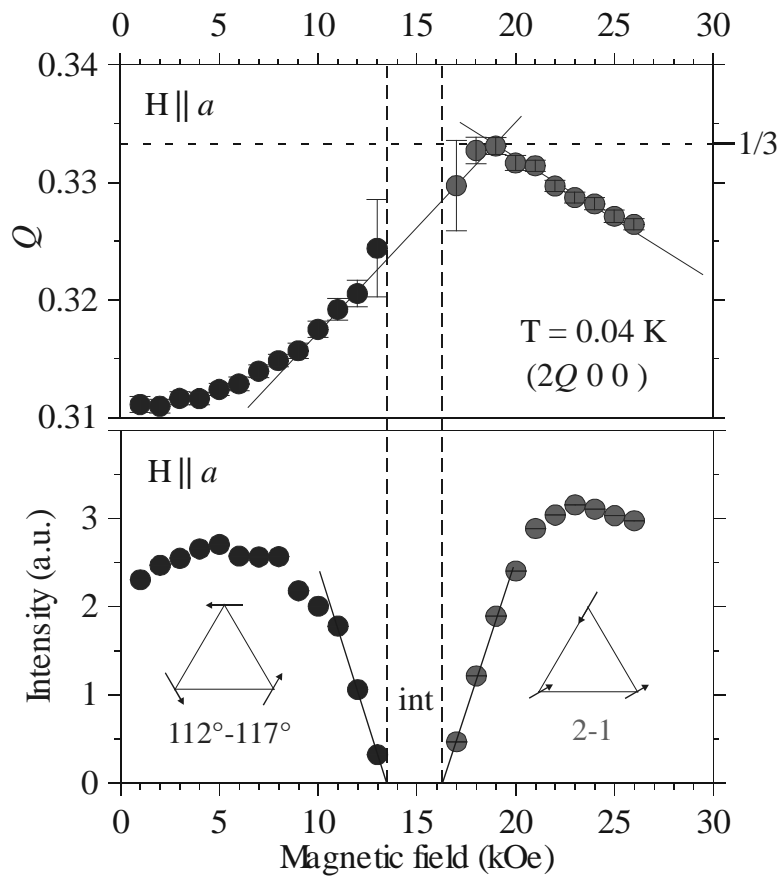

**Figure 6.** Magnetic field dependence of the ($2Q$ 0 0) reflection for $\boldsymbol{H} \parallel \boldsymbol{Q}$ measured at $T = 0.04$ K. Upper portion: field dependence of $Q$. Lower portion: field dependence of the intensity. Lines are guides to the eye. The temperature dependence of the intensity suggests the presence of three different magnetic phases.

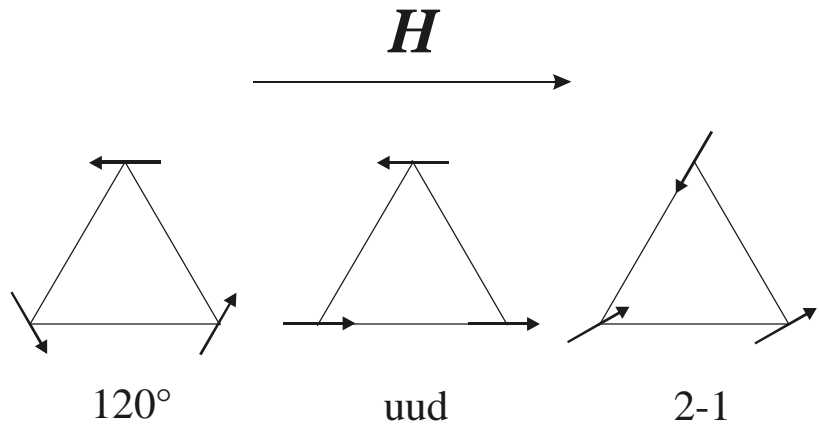

**Figure 7.** Three spin structures of the ideal frustrated triangular lattice in an external magnetic field as they occur with increasing field. The intermediate field "uud" phase has a magnetization of 1/3 and is believed to be stabilized by fluctuations.